\documentclass{neuthist18}

\usepackage{amsmath}
\usepackage{amssymb}
\newcommand{\Eps}{\varepsilon}
\newcommand{\eps}{\varepsilon}

\def\be{\begin{equation}}
\def\ee{\end{equation}}
\def\bea{\begin{eqnarray}}
\def\eea{\end{eqnarray}}

\newcommand{\Dlt}{\Delta\delta}
\newcommand{\Dmq}{\Delta m^2}
\newcommand{\dCP}{\delta_{\rm CP}}
\newcommand{\eVq}{\ensuremath{{\rm eV}^2}}

\newcommand{\Photo}{}

\begin{document}
\vspace*{4cm}
\title{Neutrino Masses and Mixing: A Little History for a Lot of Fun\footnote{
Invited Talks at the ``History of Neutrino'' Conference, September 2018, Paris}}
\author{M.C. Gonzalez-Garcia}
\address{Departament  de  Fisica  Quantica  i  Astrofisica
 and  Institut  de  Ciencies  del  Cosmos,  Universitat
 de Barcelona, Diagonal 647, E-08028 Barcelona, Spain\\
Instituci\'o Catalana de Recerca i Estudis Avancats,
Pg. Lluis  Companys  23,  08010 Barcelona, Spain.\\
C.N. Yang Institute for Theoretical Physics, Stony Brook University,
Stony Brook NY11794-3849,  USA}
\maketitle
\abstracts{In this talk I present my personal summary of the progress
  on the determination of the masses of the neutrinos and of the
  leptonic flavour mixing from the combined analysis of the experimental
  results.}
\section{Prologue: Perspective from Fall 2018}
\label{sec:present}
I start this talk on the piece of history that the organizers
asked me to cover by describing the end of the history so far, ie
the present, September 2018.

We stand here 50 years after the first results from the Chlorine
experiment which measured a flux of $\nu_e$ from the Sun which
came out to be a bit too small compared to the expectations \cite{kirsten}
launching the neutrino flavour oscillation adventure.
Since then we have been gathering data from a large number of neutrino
experiments performed with a variety of neutrino sources, and covering a wide
range of neutrino energies. The progress (which sometimes came with the 
associated confusion) on the experimental front has been
covered in the talks of (in order of appearance) T. Kirsten~\cite{kirsten},
P. Lipari~\cite{paolo}, J. Learned~\cite{learned}, P. Vogel~\cite{vogel},
K. Kleinkecht~\cite{lbl1}, G. Feldman~\cite{feldman}, T. Kajita~\cite{kajita},
A. McDonald~\cite{art}, and T. Lasierre~\cite{reactor2}. From their results 
we have established with high or at least good precision that: 

$\bullet$ Atmospheric $\nu_\mu$  and
$\bar\nu_\mu$ disappear most likely converting to $\nu_\tau$
and $\bar\nu_\tau$.
The results show an energy and distance dependence perfectly described by 
mass-induced oscillations. 

$\bullet$ Accelerator $\nu_\mu$  and
$\bar\nu_\mu$ disappear over distances of $\sim$ 200 to 700 Km. 
The energy spectrum of the results show a clear oscillatory behaviour
also in accordance with mass-induced oscillations. 

$\bullet$  Solar $\nu_e$ convert to $\nu_{\mu}$ and/or $\nu_\tau$.
The observed energy dependence of the effect is well described by  
neutrino conversion in the Sun matter according to the MSW effect
\cite{msw}. 

$\bullet$ Reactor $\bar\nu_e$ disappear over distances of $\sim$ 200 Km
and $\sim$ 1.5 km with different probabilities.
The observed energy spectra show two different mass-induced
oscillation wavelengths:
at short distances in agreement with
the one observed in accelerator $\nu_\mu$  disappearance,
and a long distance compatible with the required
parameters for  MSW conversion in the Sun.

$\bullet$ Accelerator $\nu_\mu$ and $\bar\nu_\mu$  appear as $\nu_e$
and $\bar\nu_e$ at distances  $\sim$ 200 to 700 Km. 

All these results imply that {\sl neutrinos are massive} and {\sl there is 
physics beyond the Standard Model (SM)}.  
The logic behind this statement is that a fermion mass term couples
right-handed and left-handed fermions. But the SM, a gauge theory based
on the gauge symmetry $SU(3)_{\rm C}\times SU(2)_{\rm L}\times
U(1)_{\rm Y}$ -- spontaneously broken to $SU(3)_{\rm C}\times
U(1)_{\rm EM}$ by the the vacuum expectation value of a Higgs doublet
field $\phi$ --, contains three fermion generations which reside
in the chiral representations of the gauge group
{\sl required} to describe their interactions.
As such, right-handed fields are included for charged fermions
since they are needed to build the
electromagnetic and strong currents.  But no right-handed neutrino is
included in the model because neutrinos are neutral and colourless and
therefore the right-handed neutrinos are singlets 
of the SM group (hence unrequired).
This also implies that total lepton number ($L$) is a global a symmetry of
the model. A symmetry which is non-anomalous. 
So within the framework of the SM no mass term can be built for the neutrinos 
at any order in perturbation theory neither from non-perturbative
effects. This is,  SM predicts that neutrinos are {\sl strictly} 
massless.  Consequently, there is neither mixing nor CP violation in the
leptonic sector. Clearly this is in contradiction with the neutrino
data as summarized above.

The fundamental question opened by those results is that of the
underlying  beyond the standard model
theory for neutrino masses and P. Ramond~\cite{ramond}
has discussed such theoretical implications.
But as for the description of the data
we can live with an effective model consisting of 
the Standard Model minimally extended to include neutrino masses. 
This minimal extension is what I call  {\sl The New Minimal Standard Model}
(NMSM). 

The two minimal extensions to give neutrino mass and explain the data are:

$\bullet$ Introduce $\nu_R$  and impose $L$ conservation so after spontaneuous
electroweak symmetry breaking
\begin{equation}
{\cal L}_D={\cal L}_{SM}-M_\nu \bar \nu_L \nu_R +h.c. 
\label{eq:dirac}
\end{equation}
In this case mass eigenstate neutrinos are Dirac fermions, 
ie $\nu^C\neq \nu$.

$\bullet$  Construct a mass term only with the SM left-handed neutrinos
by allowing $L$ violation 
\begin{equation}
{\cal L}_M={\cal L}_{SM}-\frac{1}{2}M_\nu \bar \nu_L \nu^c_L +h.c. 
\end{equation}
In this case the mass eigenstates are Majorana fermions, $\nu^C=\nu$.
Furthermore the Majorana mass term above also breaks the electroweak 
gauge invariance. In this respect ${\cal L}_M$ 
can only be understood as a low energy 
limit of a complete theory while ${\cal L}_D$ is formally self-consistent.

Either way, in the NMSM flavour is mixed in the CC
interactions of the leptons, and a leptonic mixing matrix appears
analogous to the CKM matrix for the quarks.  However the discussion of
leptonic mixing is complicated by two factors. First the number
massive neutrinos is unknown, since there are no constraints on
the number of right-handed, SM-singlet, neutrinos. Second, since
neutrinos carry neither color nor electromagnetic charge, they could
be Majorana fermions. As a consequence the number of new parameters in
the model depends on the number of massive neutrino states and on
whether they are Dirac or Majorana particles.

In general, if we denote the neutrino mass eigenstates by $\nu_i$,
$i=1,2,\ldots,n$, and the charged lepton mass eigenstates 
by $l_i=(e,\mu,\tau)$, in the mass basis, leptonic CC interactions 
are given by
\begin{equation}
-{\cal L}_{\rm CC}={g\over\sqrt{2}}\, \overline{{l_i}_L} 
\, \gamma^\mu \, U^{ij}\,  \nu_j \; W_\mu^+ +{\rm h.c.}.
\label{CClepmas}  
\end{equation} 
Here $U$ is a $3\times n$ matrix which verifies 
$U U^\dagger= I_{3\times3}$  but in general 
$U^\dagger U \neq  I_{n \times n}$. 

Assuming only three massive states, $U$ is a $3\times 3$ matrix
which for Majorana (Dirac) neutrinos depends on 
six (four) independent parameters: three mixing angles and three (one) phases
\begin{equation}
    U =\left(\begin{array}{ccc}
	1 & 0 & 0 \\
	0 & c_{23}  & {s_{23}} \\
	0 & -s_{23} & {c_{23}}
    \end{array}\right)
    \cdot
    \left(\begin{array}{ccc}
	c_{13} & 0 & s_{13} e^{-i\delta_{\rm CP}} \\
	0 & 1 & 0 \\
	-s_{13} e^{i\delta_{\rm CP}} & 0 & c_{13}
    \end{array}\right)
    \cdot
    \left(\begin{array}{ccc}
	c_{21} & s_{12} & 0 \\
	-s_{12} & c_{12} & 0 \\
	0 & 0 & 1
    \end{array}\right)
    \cdot
    \left(\begin{array}{ccc}
	e^{i \eta_1} & 0 & 0 \\
	0 & e^{i \eta_2} & 0 \\
	0 & 0 & 1
    \end{array}\right),
    \label{eq:U3m}
\end{equation}
where $c_{ij} \equiv \cos\theta_{ij}$ and $s_{ij} \equiv
\sin\theta_{ij}$.  In addition to the Dirac-type phase
$\delta_{\rm CP}$, analogous to that of the quark sector, there are
two physical phases $\eta_i$ associated to the Majorana character of neutrinos.

A consequence of the presence of neutrino masses and the leptonic mixing is the
possibility of mass-induced flavour oscillations of the neutrinos as described
in the talks of S. Bilenky\cite{bilenky} and E. Akhmedov \cite{Akhmedov}.
The flavour transition probability presents an oscillatory $L$ dependence
with phases proportional to $\sim \Delta m^2 L/E$ and amplitudes proportional
to different elements of mixing matrix. 
So in what respects the information that the data give us on the
new parameters in the model, neutrino oscillations are sensitive to mass
squared differences and to the angles and phases in the mixing matrix,
but do not give us information on the absolute value
of the masses. Also the Majorana phases cancel in the oscillation
probability.

As mentioned above, the observed energy and distance dependence of the
data displays two distinctive  oscillation wavelengths.
Thus the minimum scenario requires the mixing between the three known
flavour neutrinos of the standard model. There are several 
possible conventions for the ranges of the angles and ordering of the
states. The community finally agreed to a convention in which  
the angles $\theta_{ij}$ are taken to lie in the first quadrant,
$\theta_{ij} \in [0,\pi/2]$, and the phase $\dCP \in [0, 2\pi]$.
Values of $\dCP$
different from 0 and $\pi$ imply CP violation in neutrino oscillations
in vacuum. In this convention the smallest mass splitting is taken to be
$\Delta m^2_{21}$ and it is positive by construction. 
There are two possible non-equivalent orderings for the mass 
eigenvalues: $m_1\ll m_2< m_3$ so $\Dmq_{21} \ll \Dmq_{32} (\simeq \Dmq_{31} > 0) $,
refer to as Normal ordering (NO), 
and $m_3\ll m_1< m_2$ so $\Dmq_{21} \ll -(\Dmq_{31} \simeq  \Dmq_{32} < 0)$ refer to as
Inverted ordering (IO).

In total the 3-$\nu$ oscillation analysis of the existing data involves six
parameters: 2 mass differences (one of which can be positive or negative), 
3 mixing angles, and the CP phase. I summarize in Table \ref{tab:expe}
the different experiments which  dominantly contribute
to the present determination of the different parameters in the chosen
convention.\vskip -0.2cm
\begin{table}[h]
\centering
\caption{Experiments contributing to the present determination of  
the oscillation parameters.} 
\begin{tabular}{l|l|l}
Experiment & Dominant &  Important  \\
\hline
Solar Experiments &  {$\theta_{12}$} 
&  {$\Delta m^2_{21}$}  , {$\theta_{13}$} 
\\
Reactor LBL (KamLAND)  &  
{$\Delta m^2_{21}$}  
& {$\theta_{12}$}   , {$\theta_{13}$} 
\\
Reactor MBL (Daya-Bay, Reno, D-Chooz)   
&  {$\theta_{13}$}, {$|\Delta m^2_{31,32 }|$} &
\\
Atmospheric Experiments (SK)  
&  {$\theta_{23}$}  &
{$|\Delta m^2_{31,32}|$}, 
{$\theta_{13}$},{$\delta_{\rm CP}$}
\\
Accel LBL $\nu_\mu$,$\bar\nu_\mu$,
Disapp (K2K, MINOS, T2K, NO$\nu$A) 
&  {$|\Delta m^2_{31,32 }|$}   &
{$\theta_{23}$} \\
Accel LBL $\nu_e$,$\bar\nu_e$ App (MINOS, T2K, NO$\nu$A)  
&  {$\delta_{\rm CP}$}   &  
$\theta_{13}$  ,  {$\theta_{23}$}
\end{tabular} 
\label{tab:expe}
\end{table}
\begin{figure}[h]
 \centering
  \includegraphics[width=\textwidth]{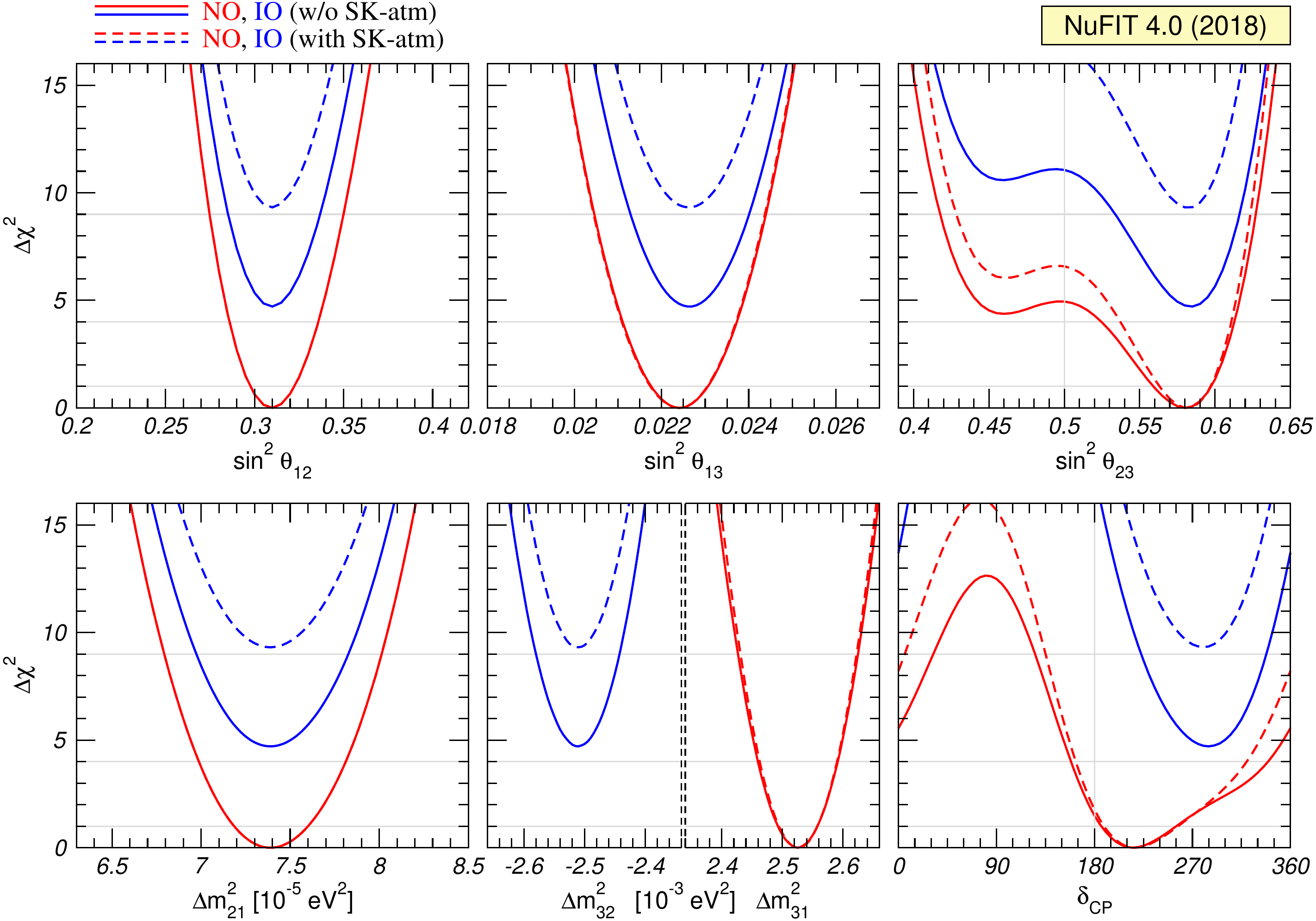}
  \caption{Global $3\nu$ oscillation analysis.  The red (blue) curves
    are for Normal (Inverted) Ordering. 
    Results for different
    assumptions concerning the analysis of data from reactor
    experiments are shown as explained in the text.}
  \label{fig:chisq-glob}
\end{figure}

The table shows that the determination of the leptonic parameters requires
global analysis of the data from the differnt experiments. Over the years
these analysis have been in the hands of a few  phenomenological groups.
The results I summarize here are from
the updated analysis in Ref.~\cite{Esteban:2018azc}
\footnote{Strictly speaking these are not the results which I presented in the talk as we were still
  making the analysis of the data presented in the summer conferences (
  and, as commented after the talk, I am not so fast anymore). But since
  the goal was to present the status at Sept 2018, I decided
  to include the results which I have now including the effect of the data
  released in the summer 18.}.
In Fig.~\ref{fig:chisq-glob} I show the determination of the six
parameters from that analysis. 

Defining the $3\sigma$ relative precision of the parameter by
$2(x^{\rm up} - x^{\rm low}) / (x^{\rm up} + x^{\rm low})$, where
$x^{\rm up}$ ($x^{\rm low}$) is the upper (lower) bound on a
parameter $x$ at the $3\sigma$ level, one reads the following
$3\sigma$ relative precision (marginalizing over ordering) :
\begin{equation}\label{eq:precision}
  \begin{array}{l@{\,,\qquad}l@{\,,\qquad}l}
    14\% \, (\theta_{12})\, &   8.9\% \, (\theta_{13})\,
    &  27\,[24]\% \, (\theta_{23})\, , \\
    16\% \,(\Dmq_{21}) \, & 7.8\, [7.6]\% \,(|\Dmq_{3\ell}|)\,
    & 100\,[92]\%\,(\dCP) \,,\\     
  \end{array}
\end{equation}
where the numbers between brackets show the impact of including
Super-Kamiokande atmospheric resutls (SK-atm)
in the precision of that parameter determination (I will comment
more on this point in Sec.~\ref{sec:23}). We notice
that as $\Delta\chi^2$ shape for $\dCP$ is clearly not gaussian this
evaluation of its ``precision'' can only be taken as indicative. 
We see that the most unclear issues are: the mass ordering discrimination,
the determination of $\sin^2\theta_{23}$, and the leptonic CP phase
$\dCP$. In brief:

$\bullet$ The best fit is for the normal mass ordering. Inverted ordering
is disfavoured with a $\Delta\chi^2 = 4.7 \, (9.3)$ without (with) SK-atm.

$\bullet$ Preference for the second octant of $\theta_{23}$,
  with the best fit point located at $\sin^2\theta_{23} = 0.58$. Values with
  $\sin^2\theta_{23} \le 0.5$ are disfavoured with $\Delta\chi^2 = 4.4
  \, (6.0)$ without (with) SK-atm.

$\bullet$ The best fit for the complex phase is at $\dCP = 215^\circ$.
  The CP conserving value of $180^\circ$, which now
  is only disfavoured with $\Delta\chi^2 = 1.5$ (1.8) without (with) SK-atm.

\section{The Main Track History: Construction of the 3$\nu$ Paradigm}
\subsection{My Prehistory: Before mid 1990's}
\label{sec:prehistory}
\begin{figure}[h]
 \centering
  \includegraphics[width=0.35\textwidth]{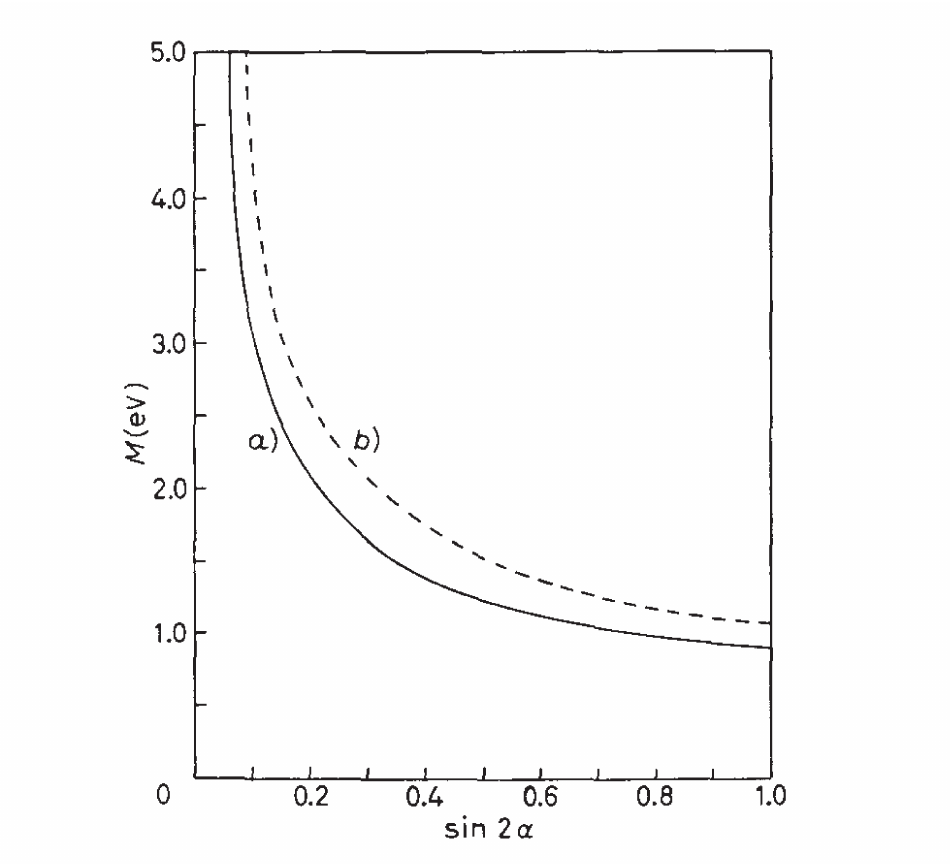}
  \caption{First oscillation parameter plot{\protect\cite{first}}? (the regions
  shown at 68\% (a) and 95\% (b) CL).}
  \label{fig:first}
\end{figure}

After describing where we are at the present, we need to decide where we
start our look to the past. The topic of my talk was the history
of the determination of neutrino properties from combined data analysis.
As for me the goal of such analysis is to provide that determination in a
statistically meaningful manner, I searched for the first time in which
neutrino flavour transition data was used in such a way, and the first
mass-mixing allowed region were presented. The first paper I found with such
a plot was Bellotti {\sl etal}~\cite{first} from 1976 which,
interpreting Gargamelle data in terms of the non observation of
$\nu_\mu\rightarrow \nu_e$ oscillation (though this was not an article signed as the
experimental collaboration), obtained an exclusion plot on some
$\Delta m^2$ and mixing angle $\alpha$ with some CL,
which I show in Fig.~\ref{fig:first} (the main difference with our present
plots is the use of the variables $M\equiv\sqrt{\Delta m^2}$ and
$\sqrt{\sin^2 2\alpha}$).

\begin{figure}[h]
 \centering
\hspace*{-1.1cm}\includegraphics[height=1.07\textwidth,angle=90]{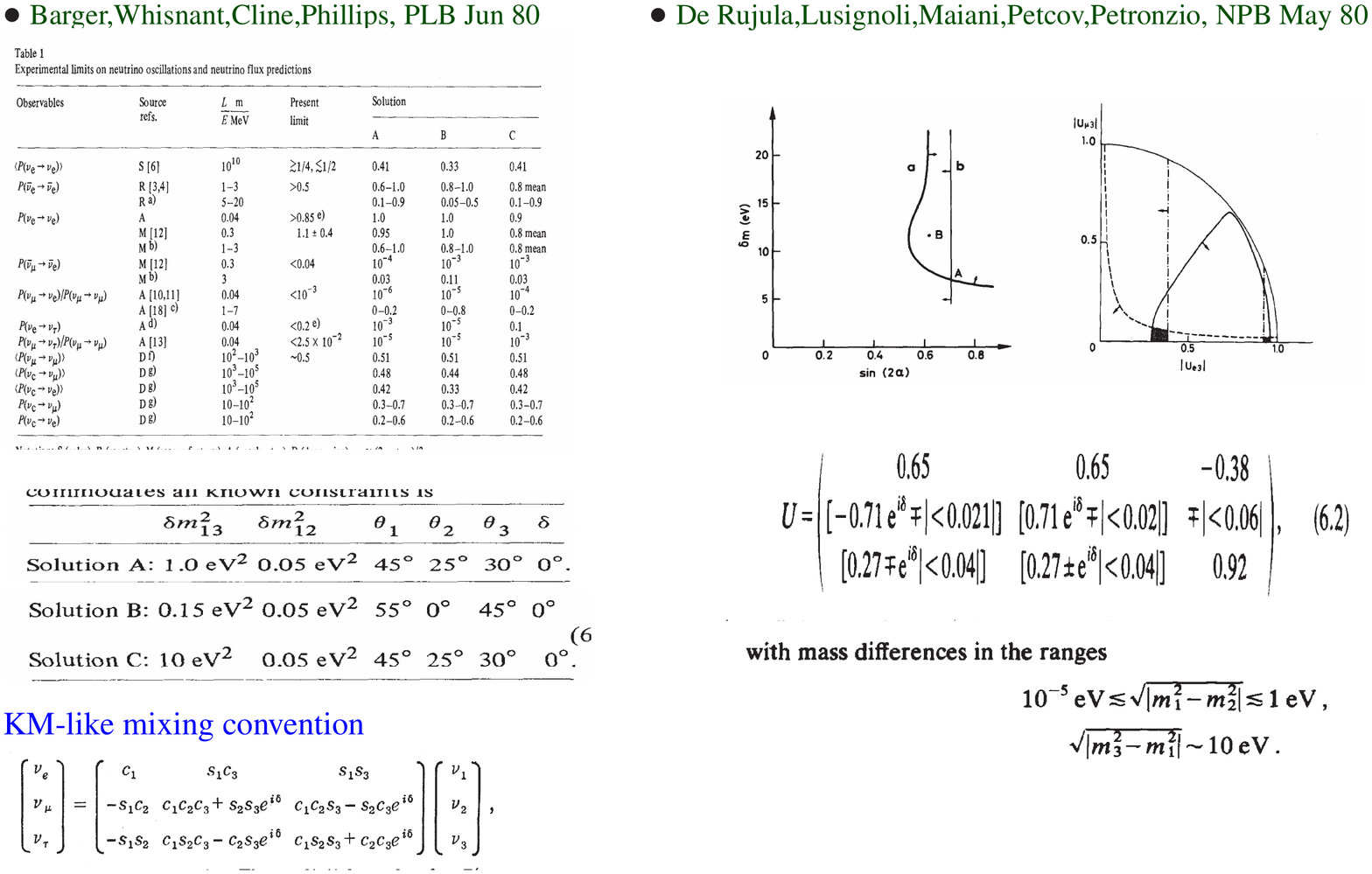}
\vglue -1cm
\caption{Examples of early {\it global} descriptions of oscillation results
  in Ref.~\protect\cite{Barger:1980ry}  (left) and
Ref.~\protect\cite{DeRujula:1979brg} (right).
}
\label{fig:pre_global}
\end{figure}

By the 80's such plots had become customary to present the results of the
reactor and neutrino fix target experiments. At that time the data was not
precise enough to allow for what I would consider {\it global analysis}
of the experimental results in the statistical sense I defined above.
But that did not prevent phenomenologist of the time to search for possible
values of neutrino parameters which could somehow describe the bulk of
experimental results.
I show in my slide in Fig.~\ref{fig:pre_global} two examples of such type
of studies from Refs.\cite{Barger:1980ry,DeRujula:1979brg}.
Besides the audacity behind these efforts, I found interesting that in both
cases one of the mass differences pointed out towards ${\cal O}$ (eV)
mass scale (see in particular the oscillation parameter region on the right). Looking at what 
experimental result was driving this, I found that already the early reactor
neutrino data was interpreted  as a {\it hint} (latter on withdraw~\cite{vogel})
of ${\cal O}$ (eV$^2$)  neutrino oscillations.
In the last years a {\it reactor neutrino anomaly} has been suggested
which points towards
the same scale and it is one of the pillars of the present $eV$ sterile neutrino
constructions which I will discuss in Sec.~\ref{sec:sterile}.
There is nothing new under the Sun.

As for what I would consider proper global/combined analysis, at the time
I entered into the field, mid 90's, the state of the art was the 3$\nu$ analysis
of Fogli and Lisi~\cite{Fogli:1993ck} which I devotedly studied as my
way of learning the subject \footnote{My motivation indeed was triggered by my
long-time collaborator JJ Gomez-Cadenas, an experimentalist working in
NOMAD, an experiment searching for $\nu_\mu\rightarrow\nu_\tau$  at short
baselines.
The early atmospheric neutrino data pointed out towards a much longer
baseline for this channel, but the LSND \cite{lsnd} result on
$\nu_\mu\rightarrow\nu_e$, which had recently made public, opened the possibility
of a high enough $\Delta m^2$ for NOMAD to see a signal if all data could be
put together. But to fit LSND
data together with the solar and atmospheric results required a fourth sterile
neutrino \cite{GomezCadenas:1995sj}. And to do this analysis I had to learn
3$\nu$ fits.}.

I was lucky enough to enter into the field right at the time
when the experimental results which established beyond doubt
mass-induced neutrino oscillations started to pour in. In what follows
I will try to illustrate the progress we made in the determination
of the neutrino parameters
as more data came in, by classifying the results by the
{\sl parameter sectors} in the 3$\nu$ oscillation framework. 

\subsection{Progress by {\it Sectors}: $\Delta m^2_{21}$ and $\theta_{12}$ }
\label{sec:12}
\begin{figure}[h]
 \centering
\hspace*{-0.6cm}  \includegraphics[height=1.15\textwidth,angle=90]{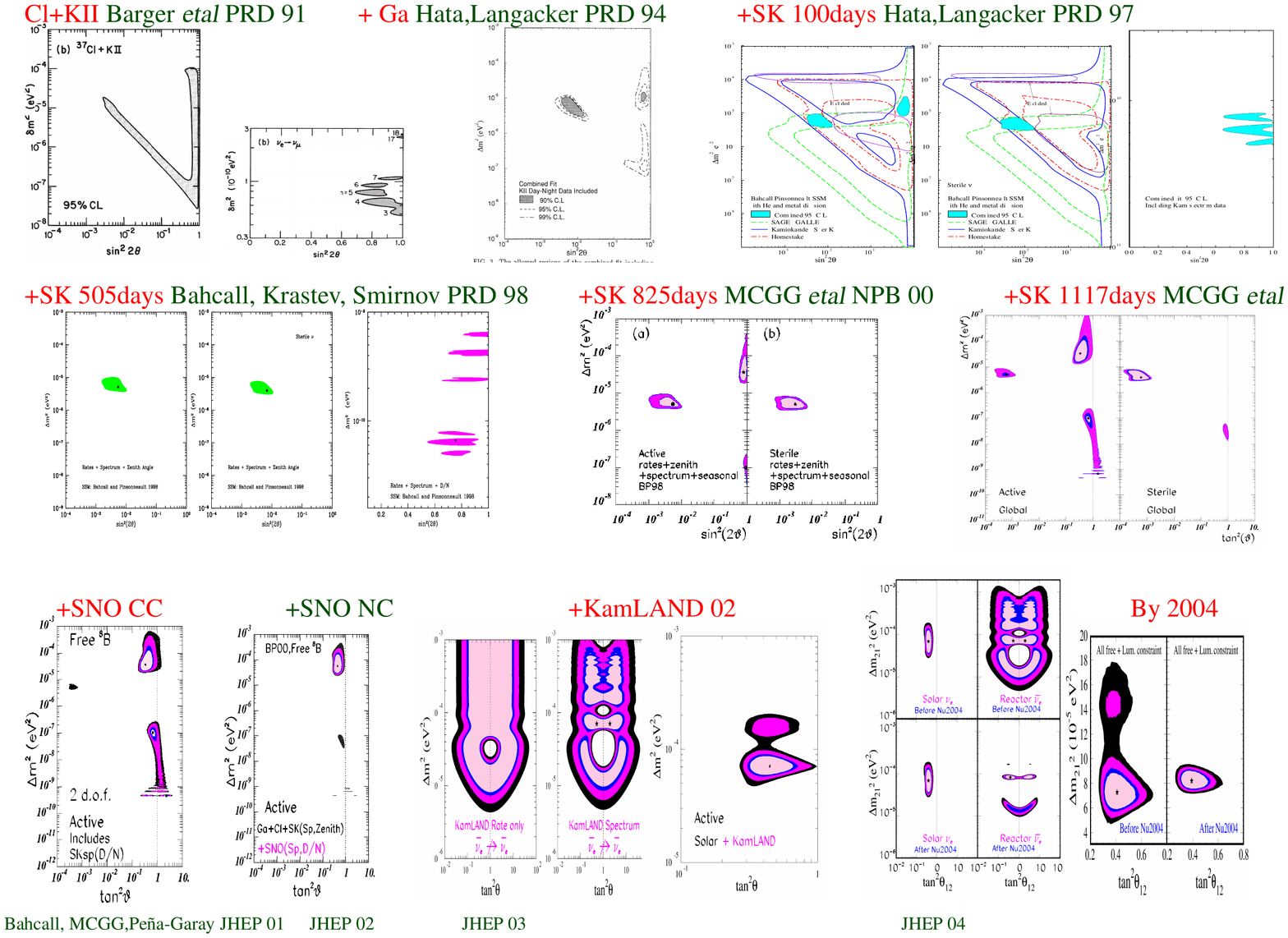}
  \caption{Slide with compilation of the parameter
    determination in the {\it solar sector}  with figures taken from
    Refs.~\protect\cite{Barger:1990qt,Hata:1993rk,Hata:1997di,Bahcall:1998jt,GonzalezGarcia:1999aj,GonzalezGarcia:2000sk,Bahcall:2001zu,Bahcall:2002hv,Bahcall:2002ij,Bahcall:2004ut}}
  \label{fig:12}
\end{figure}

As seen in table~\ref{tab:expe}, within the convention we have chosen,
$\Delta m^2_{12}$ and $\theta_{12}$ are dominantly determined by solar
neutrino experiments and KamLAND long baseline reactor data. This is
probably the sector where the historical progress in the parameter
determination is more striking. I have plotted in the slide in
Fig.~\ref{fig:12} the parameter plots in this sector presented in a selection
of consecutive references from different groups together with the
data included in each analysis\cite{Barger:1990qt,Hata:1993rk,Hata:1997di,Bahcall:1998jt,GonzalezGarcia:1999aj,GonzalezGarcia:2000sk,Bahcall:2001zu,Bahcall:2002hv,Bahcall:2002ij,Bahcall:2004ut}.

From the top row we see how the four distinct parameter
regions for $\nu_e$ oscillations into active neutrinos (any combination
of $\nu_\mu$ and $\nu_\tau$) emerged in the analysis of the solar neutrino
data at the time: small mixing angle (SMA,
with $\Delta m^2\sim 10^{-5}$ eV$^2$,
$\sin^22\theta\sim 10^{-2}$--$10^{-3}$), large mixing angle (LMA, with
$\Delta m^2\sim 10^{-4}$ eV$^2$, $\sin^22\theta\sim 0.5$--$1$),
low mass (LOW with $\Delta m^2\sim 10^{-7}$ eV$^2$, $\sin^22\theta\sim 1$)
and vacuum (or just-so, with $\Delta m^2\sim 10^{-10}$ eV$^2$,
$\sin^22\theta\sim 0.5$--$1$).
Oscillations into pure sterile neutrinos were also considered. The modified
matter potential for $\nu_e\rightarrow\nu_s$ implied that 
they only could lead to a good global description of the solar data
with SMA parameters.
With the arrival of Super-Kamiokande day-night and spectral data
(see second row)  the situation became a bit unclear for the first two years
as first  SMA seemed favoured but soon latter LMA started giving a better fit,
more and more so as more statistics was accumulated.
In the third row we see how SNO, first CC and then NC data --
besides establishing in a total model independent way the solar neutrino flavour
transition -- when included in the global analysis definitively
disfavoured SMA below 3$\sigma$ (and also $\nu_e$ oscillations into pure
sterile states) allowing only for some small LOW and
quasi-vacuum regions at that CL besides LMA. Along then came the first
results from the long baseline reactor experiment KamLAND, and, as seen in
the last plot, by 2004 a unique allowed range for these two parameters was
well established. 

Since 2004 the improvement in the determination  of $\Delta m^2_{21}$
and $\theta_{12}$  has been comparatively modest.
Historically, however the comparison of solar and KamLAND
data also played a role as giving the first hint towards a non-zero value of
$\theta_{13}$~\cite{foglihints} as I will discuss next.

\subsection{Progress by {\it Sectors}: $\theta_{13}$ }
\label{sec:13}
\begin{figure}[h]\vglue -0.5cm
 \centering
\hspace*{-0.7cm}  \includegraphics[height=1.2\textwidth,angle=90]{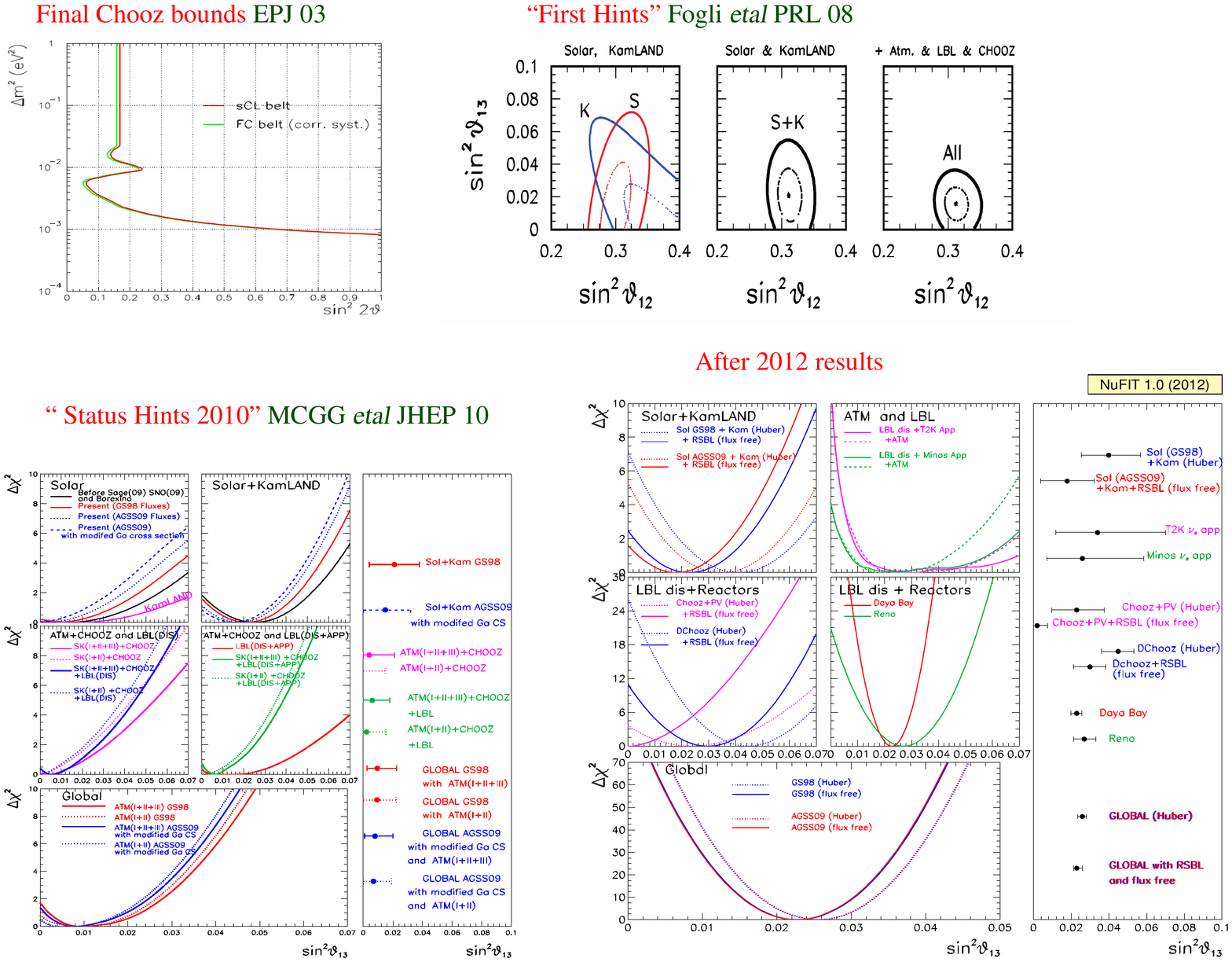}\vglue -0.3cm
  \caption{Slide illustrating the the determination of
    $\theta_{13}$ with figures taken from Refs.~\protect\cite{Apollonio:2002gd,foglihints,GonzalezGarcia:2010er,GonzalezGarcia:2012sz}.}
  \label{fig:13}
\end{figure}

I have compiled in the slide in Fig.~\ref{fig:13} some plots illustrating the
time evolution of the determination of $\theta_{13}$.

For years our most precise information on $\theta_{13}$ was the upper bound
derived from the non-observation of reactor $\bar\nu_e$ disappearance
at short distances.
The stringiest bound, shown in the first panel of that figure was provided
by the CHOOZ experiment \cite{Apollonio:2002gd}. Within their precision the best fit corresponded
to $\theta_{13}$=0. With the known hierarchy between the  oscillation
wavelengths, setting $\theta_{13}=0$ allowed for the simplification
of the 3$\nu$ analysis. For example the survival probability of solar
and KamLAND neutrinos in the framework of three neutrino oscillations
can be written as:
\begin{equation}
  \label{eq:ps3}
  P^{3\nu}_{ee} = \sin^4\theta_{13} + \cos^4\theta_{13}
  P^{2\nu}_{ee}(\Dmq_{21},\theta_{12}) \,,
\end{equation}
where we have used the fact that $L^\text{osc}_{31} = 4\pi E /
\Dmq_{31}$ is much shorter than the distance traveled by ether Solar
or KamLAND neutrinos, and  for solar neutrinos 
$P^{2\nu}_{ee}(\Dmq_{21},\theta_{12})$ should be calculated taking
into account the evolution in an effective matter density
$n^\text{eff}_{e} = n_e \cos^2 \theta_{13}$.
So for $\theta_{13}=0$ the results obtained within the 3$\nu$ mixing and
2$\nu$ mixing were exactly the same. 

However with the more precise data from both solar and KamLAND experiments,
the results obtained within the framework of 2$\nu$ oscillation
started showing some mismatch  between the best fit
value of $\theta_{12}$ in solar analysis vs the one obtained in KamLAND which
preferred a somewhat larger value. Agreement could be restored with
a non-zero value of $\theta_{13}$ because $P^{2\nu}_{ee}(\Dmq_{21},\theta_{12})$
presents the following asymptotic behaviors
\begin{align}
  \label{eq:ps2l}
  P^{2\nu}_{ee}(\Dmq_{21},\theta_{12})
  &\simeq 1 - \frac{1}{2} \sin^2(2\theta_{12})
  & {\rm for\; solar\; with}\quad
  E_\nu &\lesssim {\rm few} \times 100~{\rm KeV}
  \nonumber \\
  P^{2\nu}_{ee}(\Dmq_{21},\theta_{12})
  &\simeq \sin^2(\theta_{12})
  & {\rm for\; solar\; with }\quad
  E_\nu &\gtrsim {\rm few} \times 1~{\rm MeV}
  \nonumber \\
  P_{ee}^{2\nu}(\Dmq_{21},\theta_{12})& = 1 - \frac{1}{2}\sin^2(2\theta_{12})
  \sin^2\frac{\Dmq_{21} L}{2E} 
  & {\rm for\; KamLAND}\quad  \:. \nonumber 
\end{align}
So to obtain the same survival probability with a non-zero value of
$\theta_{13}$ at KamLAND $\theta_{12}$ should shift to lower values while
the solar region however remains pretty much at the same values of
$\theta_{12}$. This is illustrated in the triptych
on the upper right of Fig.~\ref{fig:13} taken from Ref.~\cite{foglihints}.
In our 2010 analysis~\cite{GonzalezGarcia:2010er} we found
that the effect was, however, not very
statistically
significant as seen in the compilation of the determination of $\theta_{13}$
in the lower left panels of Fig.~\ref{fig:13}.

The situation became totally clear by 2012 with the results from
T2K and specially from the medium baseline reactor experiments, Daya-Bay, Reno and Double-Chooz. In these experiments the dominant oscillation
has wavelength determined by $|\Delta m^2_{31}|$ and amplitude
$\sin^2(2\theta_{13})$ (see Eq.~(\ref{eq:mee})).
As seen in the panels in the lower right of Fig.~\ref{fig:13}~\cite{GonzalezGarcia:2012sz} in less than one
year of data from dedicated experiments the determination of a non-zero
$\theta_{13}$ was an {\sl uncontroversial}  $\sim 10\sigma$ effect. 

\subsection{Progress by {\it Sectors}: $\Delta m^2_{23}$ and $\theta_{32}$ }
\label{sec:23}
\begin{figure}[h]
\includegraphics[width=.5\textwidth]{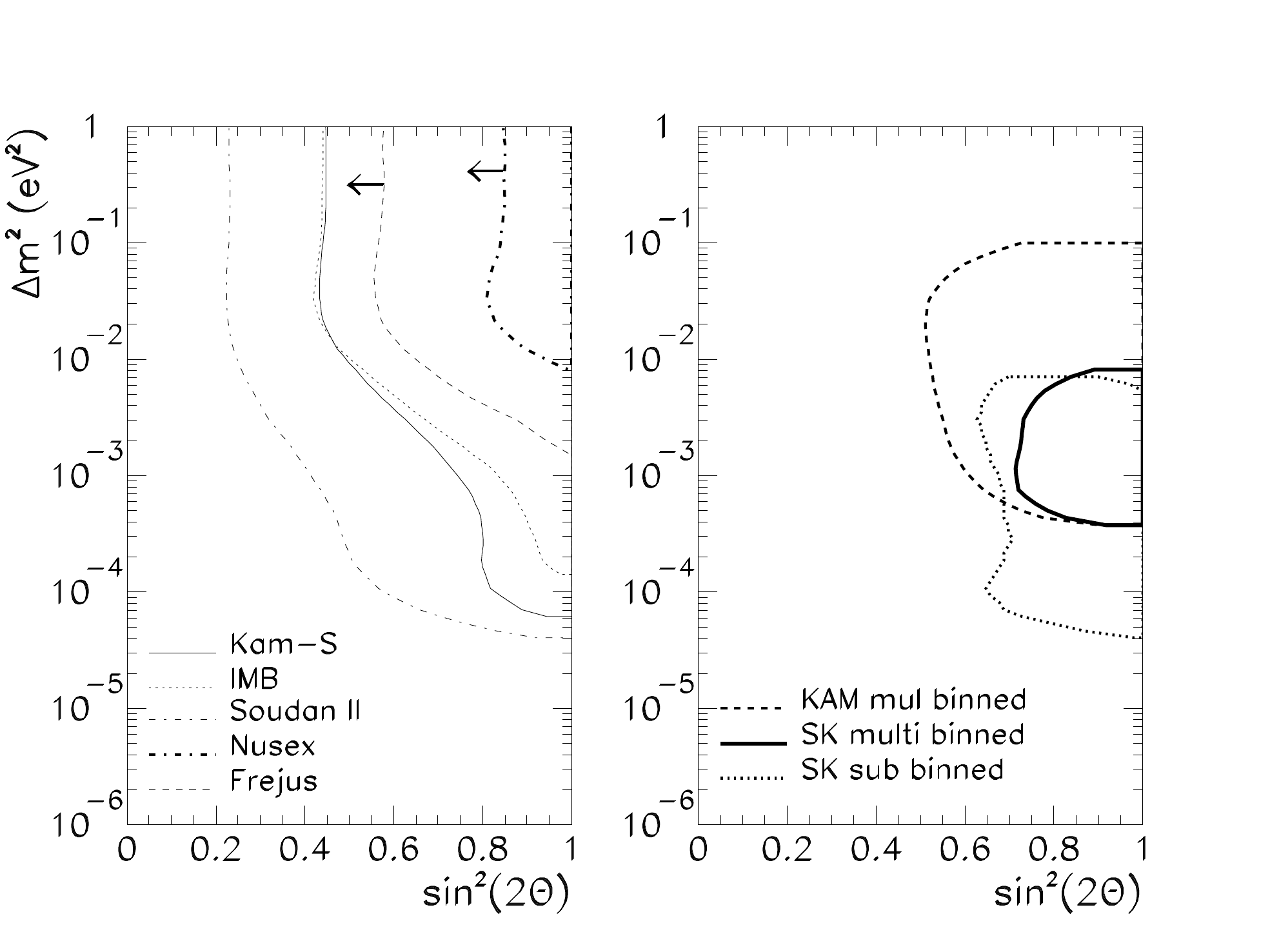}
\includegraphics[width=0.45\textwidth]{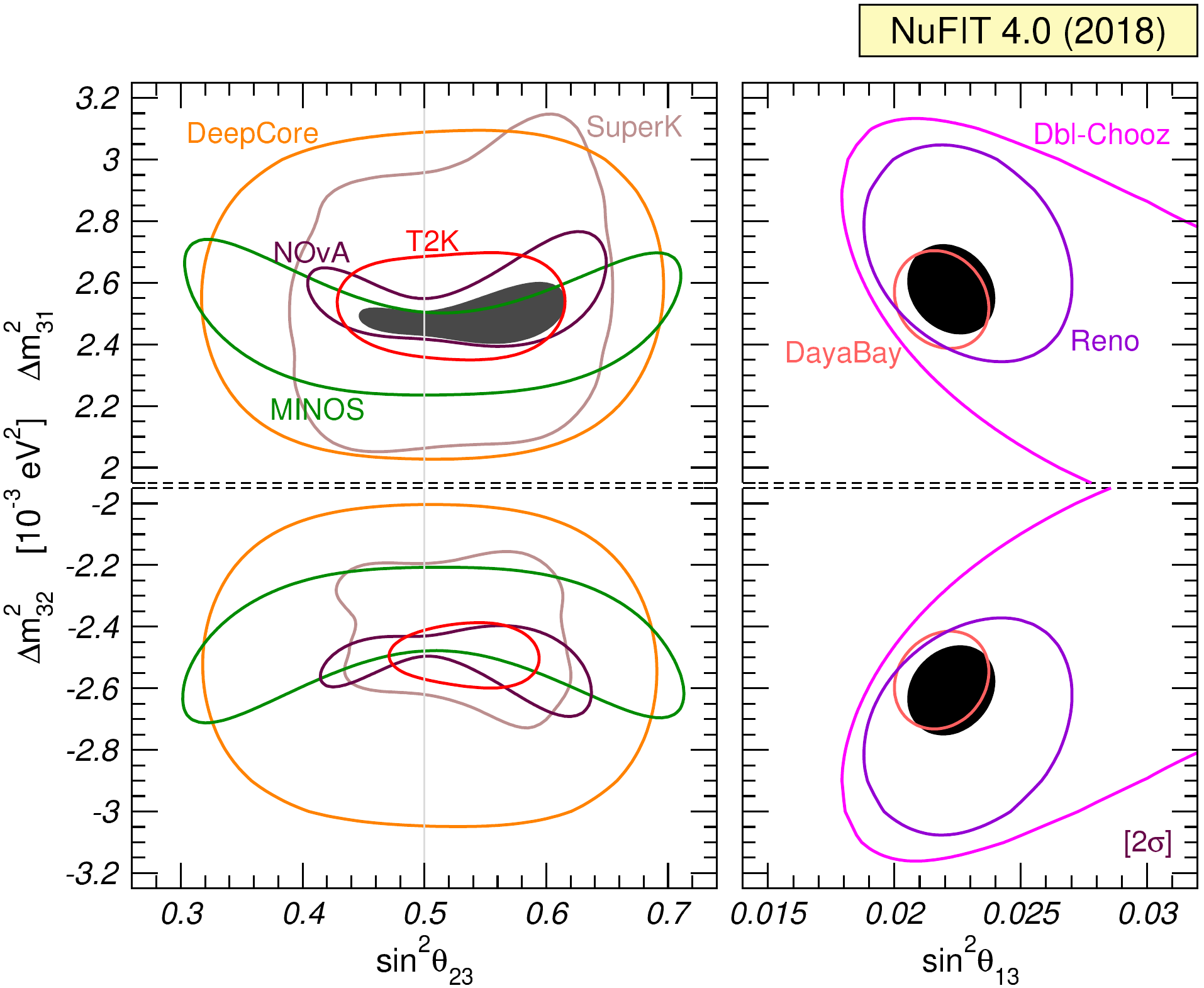}\\
\hspace*{1cm}{\footnotesize M.C G-G {\sl etal} PRD Jun 98}
  \caption{Slide illustrating the parameter determination in
    the {\it atmospheric} sector.}
  \label{fig:23}
\end{figure}

As seen in table~\ref{tab:expe}, within the convention we have chosen,
$|\Delta m^2_{23}|$ and $\theta_{23}$  are dominantly determined at present
by a combination of atmospheric, LBL and most recently the MBL reactor  
experiments. I have illustrated in Fig.~\ref{fig:23} how the allowed
regions for these parameters have changed in the last 20 years.

The year 1998 holds a special historical significance for neutrino
oscillation physics
as it was the year in which Super-Kamiokande presented the first evidence
of zenith (and therefore distance) dependence of atmospheric multi-GeV $\nu_\mu$ disappearance~\cite{kajita}.
From the point of view of parameter determination, already
with the data at the time it was possible to rule out
$\nu_\mu\rightarrow \nu_e$ as the dominant oscillation channel
for $\nu_\mu$ disappearance because its corresponding amplitude was
determined by the $\theta_{13}$ angle which was already constrained
to be too small by CHOOZ. The angular dependence
of the event rates also  disfavoured oscillations
into sterile neutrinos for which matter effects yield a flatter zenith
angle dependence. Consequently $\nu_\mu\rightarrow \nu_\tau$ was established
as the dominant flavour transition channel observed in atmospheric oscillations.
The relevant survival probability takes the form
\begin{equation}
  P_{\mu\mu}\simeq 1-(c_{13}^4 { \sin^2 2\theta_{23}}+s_{23}^2\sin^22\theta_{13})
  \sin^2\left(\frac{{ \Delta m^2_{31}} L }{4 E}\right)+
 {\cal O}(\Delta m^2_{21})\, .
  \label{eq:pmm}
\end{equation}
Therefore in the limit $\theta_{13}=0$ and $\Delta m^2_{21}=0$ the atmospheric
data analysis  determines $|\Delta m^2_{31}|=|\Delta m^2_{32}|$
and $\sin^2(2\theta_{23})$ as shown in the left panels in Fig.~\ref{fig:23}.
Experiments on the most left panel did not provided zenith angle
dependence information and therefore the allowed region
extended to arbitrary large $\Delta m^2$. At the time also some
experiments reported an effect while others did not \cite{paolo,learned}.

In the early years of this century, long baseline accelerator experiments,
starting with K2K and MINOS confirmed this picture.
Furthermore the analysis of their $\nu_\mu$ disappearance energy spectrum
provided us with the most precise determination of the mass splitting.
Precision now is in hands of T2K and NO$\nu$A as seen in Fig.~\ref{fig:13}.

The latest contribution to the determination of this mass splitting
has come in the last five years from the analysis of the spectrum of
$\bar\nu_e$ disappearance in MBL reactor experiments.
The relevant
survival probability can be approximated as
\begin{equation}
P_{ee}\simeq 1-{ \sin^2 2\theta_{13}}
\sin^2\left(\frac{{ \Delta m^2_{ee}} L}{4E}\right)-c_{13}^4\sin^2 2\theta_{12} \sin^2\left(\frac{\Delta m^2_{21} L}{4E}\right)\;\;
\label{eq:mee}
\end{equation}
with $\Delta m^2_{ee}\simeq |{\Delta m^2_{32}}|
\pm { c_{12}^2 \Delta m^2_{21}}
\simeq |{\Delta m^2_{32}}|$.
As seen in Fig.~\ref{fig:13} the precision attainable on the mass
splitting from the analysis of $\bar\nu_e$ disappearance spectrum
at MBL reactor experiments is at present comparable from that of
$\nu_\mu$ disappearance at LBL accelerator experiments.

In what respects the determination  of $\theta_{23}$, till recently
it was dominated by the analysis of SK atmospheric neutrinos and it
favoured maximal mixing. This changed with the increase precision
of the LBL experiments though in not a totally consistent direction.
The status on the maximality of $\theta_{23}$, or on the octact preference
in case of not maximality, has varied over the last years as more data was
gathered. This is still an unsettled issue.

\subsection{Ordering and $\delta_{\rm CP}$}
There is not much {\sl history} on the determination of the mass
ordering and the CP phase. It is being written as I type these
proceedings. The measurement of a not-too-small $\theta_{13}$  made it
possible to obtain some statistical significance on both from
the analysis of $\nu_e$ and $\bar\nu_e$ appearance in the present
LBL experiments, T2K and NO$\nu$A. The quest is on.

An additional issue which has come out over the recent years in this
respect, is that of how to include in the global analysis the results of
SK-atm on these effects. With the phenomenological
tools developed to analyze the data and obtain the results on the dominant
effects described above (ie on $\theta_{23}$ and $|\Delta m^2_{31}|$),
very limited sensitivity to the $\theta_{13}$, the ordering and
to $\delta_{\rm CP}$ is found. But the collaboration has developed a
more sophisticated analysis method with the aim of constructing
enriched samples which are most sensitive to these subdominant effects, 
and which cannot be technically reproduced outside of the collaboration.
Super-Kamiokande has published the results of that analysis in
the form of a tabulated $\chi^2$ map as a function of the four
relevant parameters $\Dmq_{3\ell}, \theta_{23},\theta_{13}$, and
$\dCP$. At the moment this is what is being {\it blindly} added in the
combined phenomenological analysis. As seen in Fig.~\ref{fig:chisq-glob} this
addition has a non-negligible impact on the statistical discrimination
between orderings (and somewhat less on the determination of $\delta_{\rm CP}$).

\subsection{The Neutrino Mass Scale}
\label{sec:mscale}
Oscillation experiments provide information on $\Delta m^2_{ij}$, 
and on the leptonic mixing angles, $U_{ij}$. But they are insensitive to 
the absolute mass scale for the neutrinos. 
Of course, the results of an oscillation experiment do provide a lower bound
on the heavier mass in $\Delta m^2_{ij}$, $|m_i|\geq\sqrt{\Delta m^2_{ij}}$ for
$\Delta m^2_{ij}>0$. But there is no upper bound on this mass. In particular,
the corresponding neutrinos could be approximately degenerate at a mass
scale that is much higher than $\sqrt{\Delta m^2_{ij}}$. 
Moreover, there is neither upper nor lower bound on the lighter mass $m_j$.

The only model independent information on the neutrino masses, rather than mass
differences, can be extracted from kinematic studies of reactions in which a
neutrino or an 
anti-neutrino is involved. Historically these bounds were labeled as limits on
the mass of the flavour neutrino states corresponding to the charged
flavour involved in the decay:\\[+0.2cm]
\begin{tabular}{ll}
  \hspace*{2cm}  $m_{\nu_e}\leq 2.2\,{\rm eV}$
  & From ${\rm ^3H \rightarrow\ ^3He + e^-+\overline\nu_e}$~\cite{tritium}
  \\[+0.2cm]
\hspace*{2cm} $m_{\nu_\mu}\leq 0.19\,{\rm MeV}$ \; &From 
  ${\rm\pi \rightarrow \mu + \nu_\mu}$~\cite{pdg}
 \\[+0.2cm]
\hspace*{2cm} $m_{\nu_\tau}\leq 10.2\,{\rm MeV}$& From  ${\rm \tau \rightarrow {\rm N}\pi's +\nu_\tau}$~\cite{pdg}
\end{tabular}
\\

In the presence of mixing the bounded combinations are indeed
\begin{equation}
  m_{\nu_\alpha}^2=\sum_i |U_{\alpha i}|^2 m_i^2\;,
\label{eq:mbeta}
\end{equation}
so with the values
known of the mixing matrix elements the most relevant constraint comes
from Tritium beta decay and it has been standing at the value of 2.2
eV for almost two decades. It is expected to be superseded by KATRIN
which will improve the sensitivity by about one order of magnitude.

Model dependent information on neutrino masses can also 
be obtained from neutrinoless double beta decay
$(A,Z) \rightarrow (A,Z+2) + e^{-} + e^{-}$. 
This process is the most sensitive test of the Dirac vs Majorana
nature of the neutrinos. If they are Majorana particles and in the 
context of the NMSM (in which no other source of lepton number violation
is present in the model) the rate of this process is proportional to the 
{\it effective Majorana mass of $\nu_e$},
\begin{equation}
m_{ee}
=\left| \ \ \sum_i m_i U_{ei}^2 \ \ \right|
\end{equation}
which, depends also on the three CP violating phases. 
Notice that in order to induce the $2\beta0\nu$ decay, $\nu$'s must 
Majorana particles, thus if neutrinos are Dirac particles no
information on their masses can be deduced from the non-observation 
of $2\beta0\nu$ decay.
As we heard in the talk of S. Petcov \cite{petcov} at present the most stringent
bounds are  $m_{ee}\leq 0.06$--$0.4$ where the range
spans over the nuclei involved as well as  the expected uncertainty associated
with the nuclear matrix model. 

Neutrinos, like any other particles, contribute to the total energy
density of the Universe and have impact in its evolution \cite{cosmo}.
Within what we presently know of
their masses, neutrinos are relativistic through most of the evolution
of the Universe and being very weakly interacting they decoupled early in
cosmic history.  Depending on their exact
masses they can impact the CMB spectra, in particular by altering the
value of the redshift for matter-radiation equality.  More
importantly, their free streaming suppresses the growth of structures
on scales smaller than the horizon at the time when they become
non-relativistic and therefore affects the matter power spectrum which
is probed from surveys of the LSS distribution.  Within their present
precision, cosmological observations are sensitive to neutrinos mostly
via their contribution to the energy density in our Universe,
$\Omega_\nu h^2= \sum_i m_i / (94 \,{\rm eV})$.  Therefore
cosmological data mostly gives information on the sum of the neutrino
masses and has very little to say on their mixing structure and on the
ordering of the mass states.  At present the most robust bounds 
come from the analysis of Planck results 
which within the $\Lambda$-CDM model imply 
$\sum_i m_i \leq 0.17-0.74$ eV where the range includes variations
of the data sets included in the analysis. One must always keep in mind
that these bounds apply {\sl within a given cosmological model}. Variations
of the model can relax the bounds. 

\begin{figure}\centering
\includegraphics[width=0.6\textwidth]{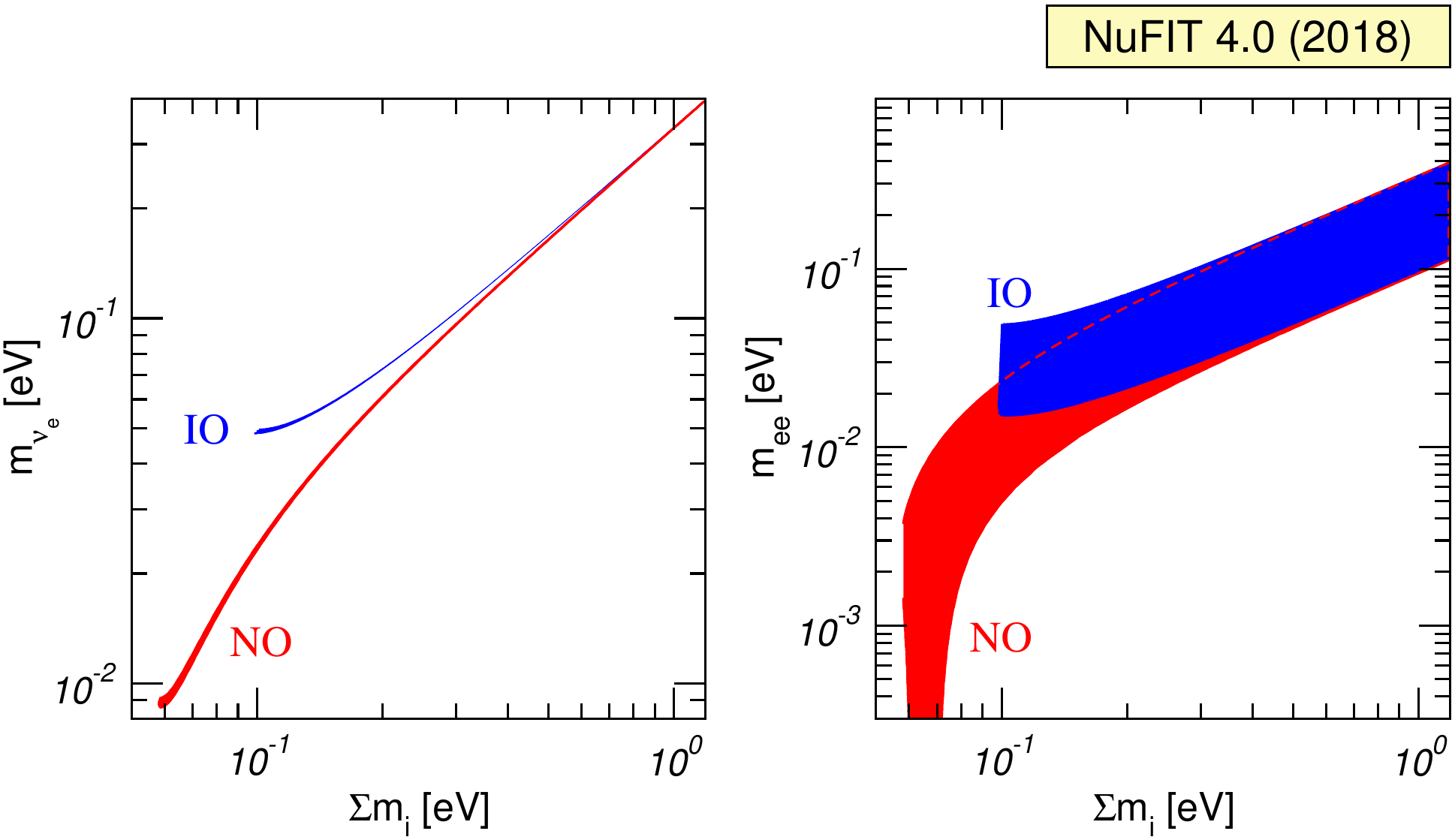}
\caption{
95\% allowed regions (for 2 dof) in the planes  
($m_{\nu_e}$,$\sum  m_\nu$) 
and ($m_{ee}$,$\sum  m_\nu$) from the global analysis of oscillation data
(full regions).}
\label{fig:mbeta} 
\end{figure}

Within the 3$\nu$ scenario, correlated information on the three probes
of neutrino masses
can be obtained by mapping the results from  the global analysis of 
oscillations presented in the previous section \cite{foglibeta}.  I show in 
Fig.~\ref{fig:mbeta} the updated status  of this exercise. The narrow
range observed in the left panel corresponds to the uncertainty associated
with the present determination of oscillation parameters, which, as seen 
in the figure, is rather small. On the contrary
the wide range observed in the right panel corresponds to the effect
of the unknown Majorana phases. From the figure one can infer that a positive
determination of two of these probes (or a sufficiently strong bound)
could help to determine the ordering of the states, and give some information
about the Majorana phases within the corresponding model assumptions. 
In this front, the quest is also ongoing with claims and disclaims on the
significance of the effects being observed.

\section{The Parallel Paths}
While the consistency of the minimal picture of mass-induced
3$\nu$ oscillations was being established, other scenarios -- either alternative
or extentended -- were proposed and as such were confronted with the data to
learn about their relevant parameters. One can consider those scenarios
as parallel paths that our history could have chosen to follow and in this
section I am going to briefly describe some of them.
\subsection{Alternative Scenarios for Flavour Conversion in Vacuum}
\label{sec:others}
Oscillations are not the only possible mechanism for neutrino flavour
transitions and over the years alternative scenarios were proposed
with nonstandard neutrino physics characterized by the presence of an
unconventional interaction (other than the neutrino mass terms) that
mixes neutrino flavours. From the point of view of neutrino
oscillation phenomenology, a critical feature of these scenarios is a
departure from the $\lambda\propto E/\Delta m^2$ dependence of the
conventional oscillation wavelength and instead $\lambda\propto
E^{-n}/\Delta\delta$ where $n$ and $\delta$ depends on the specific
mechanism. Examples include: 

$\bullet$ Violation of the equivalence principle~\cite{VEP}, due to non-
universal coupling of the neutrinos, $\gamma_1\neq \gamma_2$ to the local
gravitational potential $\phi$, or
breakdown of Lorentz invariance ~\cite{VLI,VLI1}resulting from different
asymptotic values of the velocity of the neutrinos, $c_1\neq c_2$,
for which $n=1$

$\bullet$ Non-universal coupling of the neutrinos, $k_1\neq k_2$ to a space-time
torsion field $Q$~\cite{torsion} or 
Violation of CPT resulting from Lorentz-violating effects such 
as the operator, $\bar{\nu}_L^\alpha b_\mu^{\alpha\beta} \gamma_\mu
\nu_L^\beta$, ~\cite{VLICPT1,VLICPT2,VLICPT3} which lead 
to an energy independent contribution to the oscillation wavelength.

Atmospheric neutrinos with their broad energy range and travel
distances are the ideal probe for these type of scenarios and already
with the early data from Super-Kamiokande it was possible to rule them
out as the dominant mechanism responsible for the observed flavour
transitions \cite{fogli1}.  Furthermore as data from LBL experiments
became available it was possible to constraint the subdominant
contribution from these scenarios to the standard 3$\nu$ oscillation
transitions and impose severe bounds on these extensions of the NMSM,
for example\cite{atmnp} 
\begin{equation}
\begin{array}{ll}
\Dlt= 2 |\phi|(\gamma_1- \gamma_2) 
\leq 1.6\times 10^{-24}\,, &\qquad 
{\rm for\, VEP}  \\
\Dlt = (c_1- c_2)
\leq 1.6\times 10^{-24}\,, & \qquad 
{\rm for\, VLI} \\
\Dlt= Q (k_1- k_2)
\leq 6.3\times 10^{-23}~{\rm GeV}\,, &
\qquad
{\rm for\, coupling\, to\, torsion} \\
\Dlt= b_1-b_2 \leq 5.0\times 10^{-23}~{\rm GeV}\, \,,&
\qquad 
{\rm for\,\,\,\slash\!\!\!\!\!
CPT\,,\, VLI} \;.
\end{array}
\label{eq:nplim}     
\end{equation}

\subsection{Non-standard Neutrino Interactions}
A mechanism for flavour transitions which is not fully described by the
above formalism is that of non-standard neutrino interactions (NSI) with
matter. In particular neutral current NSI's can impact the coherent
scattering of neutrinos in matter. Neutral current NSI's 
can be parametrized by effective four-fermion
operators of the form
\begin{equation}
  \label{eq:def}
  \mathcal{L}_{\rm NSI} =
  - 2\sqrt{2} G_F \Eps_{\alpha\beta}^{fP}
  (\bar\nu_{\alpha} \gamma^\mu L \nu_{\beta})
  (\bar{f} \gamma_\mu P f) \,,
\end{equation}
where $f=e,u,d$ is a charged fermion, $P=(L,R)$ and
$\Eps_{\alpha\beta}^{fP}$ are dimensionless parameters encoding the
deviation from standard interactions. These operators  contribute to the
effective matter potential in the Hamiltonian describing the evolution
of the neutrino flavour state:
\begin{equation}
  \label{eq:Hmat}
  H_\text{mat}
  = \sqrt{2} G_F N_e (x) 
  \begin{pmatrix}
    1 + \epsilon_{ee} & \epsilon_{e\mu} & \epsilon_{e\tau} \\
    \epsilon_{e\mu}^* & \epsilon_{\mu\mu} & \epsilon_{\mu\tau} \\
    \epsilon_{e\tau}^* & \epsilon_{\mu\tau}^* & \epsilon_{\tau\tau}
  \end{pmatrix}, \;\;{\rm with}\;\;
  \epsilon_{\alpha\beta}(x)
  = \sum_{f=e,u,d} \frac{N_f(x)}{ N_e(x)} \epsilon_{\alpha\beta}^{f,V}
   \,,
\end{equation}
with $N_f(x)$ being the density of
fermion $f$ along the neutrino path.  The ``1'' in the $ee$ entry in
Eq.~\eqref{eq:Hmat} corresponds to the standard MSW matter potential.
Therefore, the effective NSI parameters entering oscillations,
$\epsilon_{\alpha\beta}$, may depend on $x$ and will be generally
different for neutrinos crossing the Earth or the solar medium and as
such can be constrained by the global analysis of neutrino oscillation
data (since oscillation experiments are only sensitive to differences
between the diagonal terms in the matter potential).

The task becomes troubled by an intrinsic degeneracy in the Hamiltonian
governing neutrino oscillations which is introduced by the NSI-induced matter
potential.
In general, CPT implies that neutrino evolution is invariant if the relevant
Hamiltonian is transformed as $H \to -H^*$. In
vacuum this transformation can be realized by changing the oscillation
parameters as
\begin{equation}
  \label{eq:osc-deg}
    \Dmq_{31} \to -\Dmq_{31} + \Dmq_{21} = -\Dmq_{32} \,,\;\;\;
    \sin\theta_{12} \leftrightarrow \cos\theta_{12} \,,\;\;\;
    \delta_{\rm CP} \to \pi - \delta_{\rm CP} \,.
\end{equation}
In the standard 3$\nu$ oscillation scenario, this symmetry is broken by the
standard matter effect, and this allows for the determination of the octant
of $\theta_{12}$ and (in principle) of the
sign of $\Dmq_{31}$. However, in the presence of NSI, the symmetry can
be restored if in addition to the transformation
Eq.~\eqref{eq:osc-deg}, NSI parameters are transformed
as
\begin{equation}
  \label{eq:NSI-deg}
    (\eps_{ee} - \eps_{\mu\mu}) \to - (\eps_{ee} - \eps_{\mu\mu}) - 2 \,, \;\;\;
    (\eps_{\tau\tau} - \eps_{\mu\mu}) \to -(\eps_{\tau\tau} - \eps_{\mu\mu}) \,, \;\;\;
    \eps_{\alpha\beta} \to - \eps_{\alpha\beta}^* \qquad (\alpha \neq \beta) \,.  
\end{equation}

\begin{figure}[h]\centering
  \includegraphics[width=0.65\textwidth]{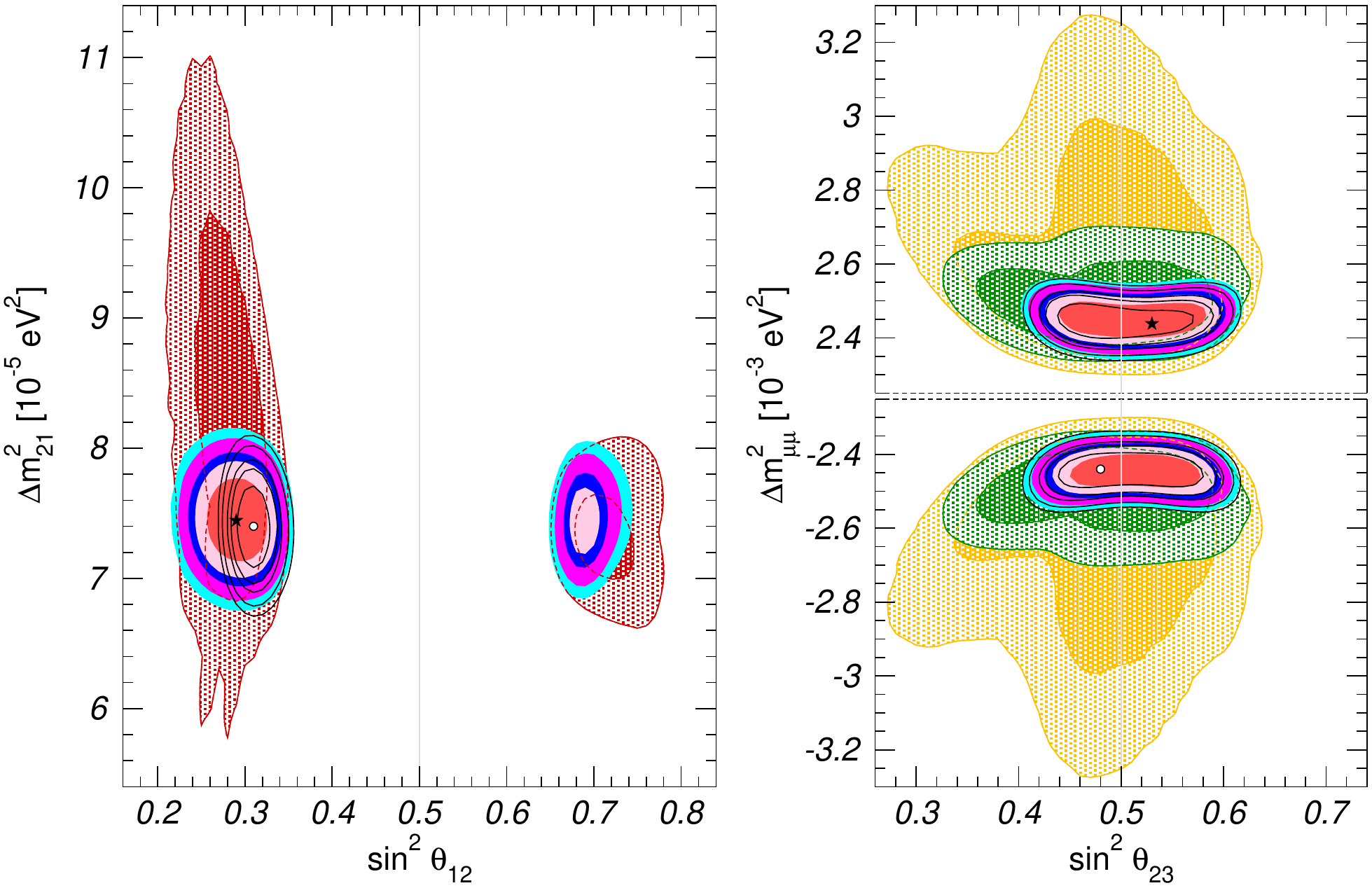}
  \caption{Two-dimensional projections of the allowed regions onto
    different vacuum parameters (on the right
    $\Dmq_{\mu\mu}\simeq \Delta m^2_{31}$) after marginalizing over the matter
    potential parameters and the undisplayed
    oscillation parameters.  The solid colored regions correspond to
    the global analysis of all oscillation data, and show the
    $1\sigma$, 90\%, $2\sigma$, 99\% and $3\sigma$ CL allowed regions;
    the best fit point is marked with a star.  The black void regions
    correspond to the analysis with the standard matter potential
    (\textit{i.e.}, without NSI) and its best fit point is marked with
    an empty dot.  For comparison, in the left panel we show in red
    the 90\% and $3\sigma$ allowed regions including only solar and
    KamLAND results, while in the right panels we show in green the
    90\% and $3\sigma$ allowed regions excluding solar and KamLAND
    data, and in yellow the corresponding ones excluding also IceCube
    and reactor data.}
  \label{fig:glb-oscil}
\end{figure}

In Fig.~\ref{fig:glb-oscil} I show the two-dimensional projections of
the allowed regions onto different sets of oscillation parameters 
from the global analysis in Ref.~\cite{Esteban:2018ppq}
in the presence of this generalized matter potential  (\ref{eq:Hmat}).
These regions are obtained after
marginalizing over the undisplayed vacuum parameters as well as the
NSI couplings. For comparison its also shown as black-contour void
regions the corresponding results with the standard matter potential,
\textit{i.e.}, in the absence of NSI.

From the figure we read the following:

$\bullet$ The determination of the oscillation parameters discussed in the
previous section is robust under the presence of NSI as large as allowed
by the oscillation data itself with the exception of the octant 
of $\theta_{12}$. This result relies on the complementarity and synergies
between the different data sets, which allows to constrain those
regions of the parameter space where cancellations between standard
and non-standard effects occur in a particular data set.   

$\bullet$ A solution with $\theta_{12}>45 ^\circ$ still provides a
good fit.  This is the {\it so-called} LMA-D solution and it was first
found in Ref.~\cite{Miranda:2004nb}. It is is a consequence of the intrinsic
degeneracy in the Hamiltonian
described above.  Eq.~\eqref{eq:osc-deg} shows that this degeneracy
implies a change in the octant of $\theta_{12}$ (as manifest in the
LMA-D). As such it cannot be ruled out by oscillation data
only. Scattering data, in particular from the finally-observed
coherent scattering in nuclei \cite{COHERENT} disfavoured it at more
then 3$\sigma$ for NSI coupling neutrinos with either up or down
quarks. But it is still allowed for more general
NSI couplings~\cite{Esteban:2018ppq}.

LMA-D requires large $\Eps_{ee}-\Eps_{\mu\mu} \sim{\cal O}(2)$ which are therefore
still allowed by the global analysis. But for all other couplings
the same global analysis sets strong constrains
on $\Eps_{\alpha\beta}$ yielding  the most restrictive
bounds on the  NSI parameters, in particular those involving $\tau$ flavour.

\subsection{Light Sterile Neutrinos}
The vast majority of the neutrino data on flavour transitions accumulated
over the years could be consistently described in the framework of three
neutrino mixing. There appeared, however, a set of anomalies
in neutrino data at relatively short-baselines (SBL) which could not.
As mentioned before, in the early 1990's LSND~\cite{lsnd}  reported the
observation of  $\nu_\mu\rightarrow \nu_e$ (
over the last decade it has been tested at MiniBooNE which also found an anomaly
though not exactly as expected from LSND \cite{miniboone}).
A few years latter it was also pointed out that the
$\nu_e$ source experiments made to test the efficiency of gallium solar
experiments did also saw a deficit compared with expectations~\cite{gallium}.
The third
set of anomalies arose in $\bar\nu_e$ reactor experiments as described in
Laserre's talk~\cite{reactor2} and came out also as a deficit compared
to theoretical expectations.
If interpreted in terms of oscillations, each of these anomalies 
points out towards a $\Delta m^2\sim {\cal O}({\rm eV}^2)$ 
and consequently cannot be described within the context of the  3$\nu$
mixing described in the previous section. They require, instead,
the addition of one or more additional neutrinos which must be
{\sl sterile}, {\it i.e.} elusive
to Standard Model interactions, to account for the constraint of the
invisible $Z$ width which limits the number of light weak-interacting
neutrinos to be $2.984 \pm 0.008$. 

\begin{figure}[t] \centering
    \includegraphics[width=0.42\textwidth]{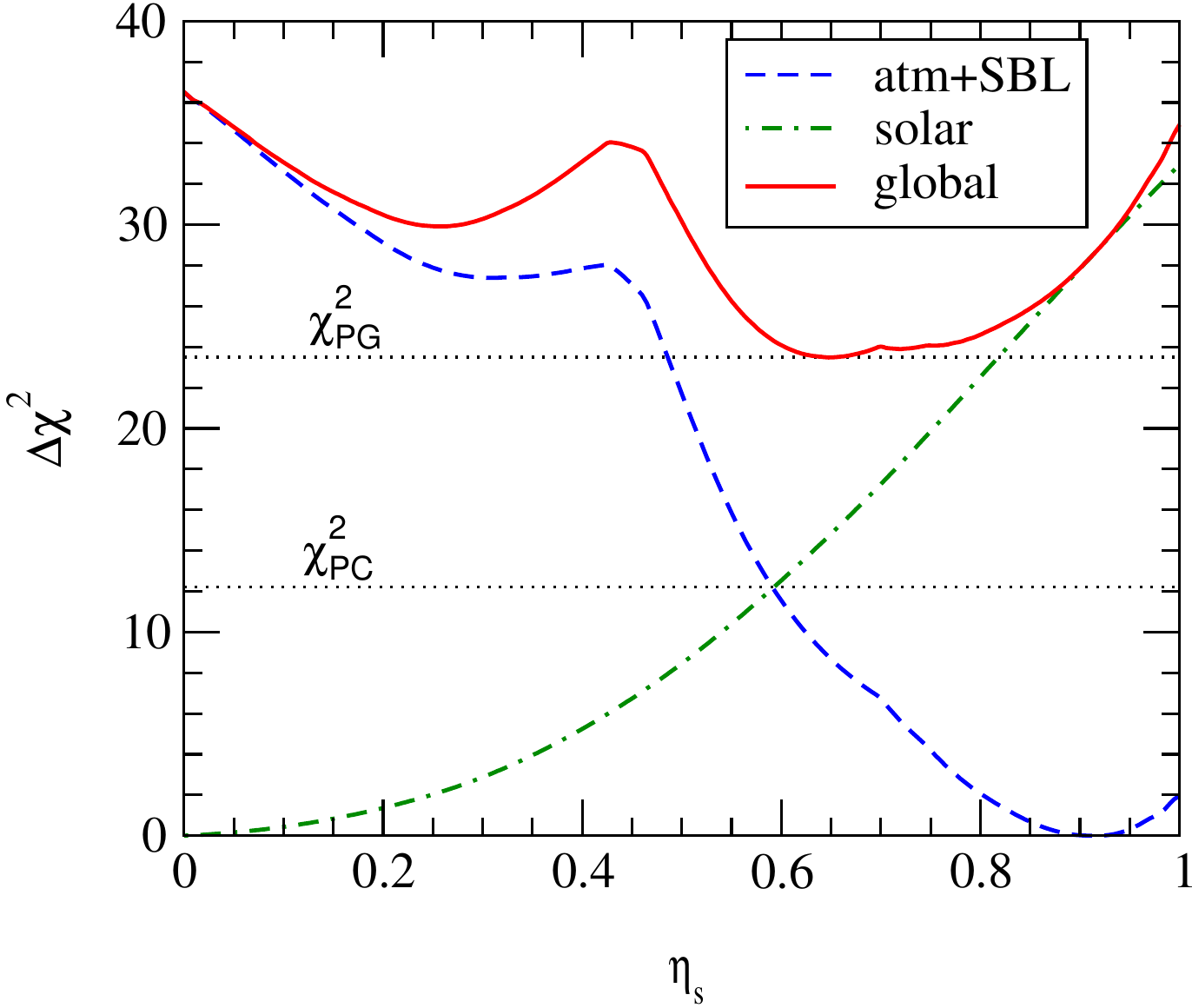}
    \includegraphics[width=0.4\textwidth]{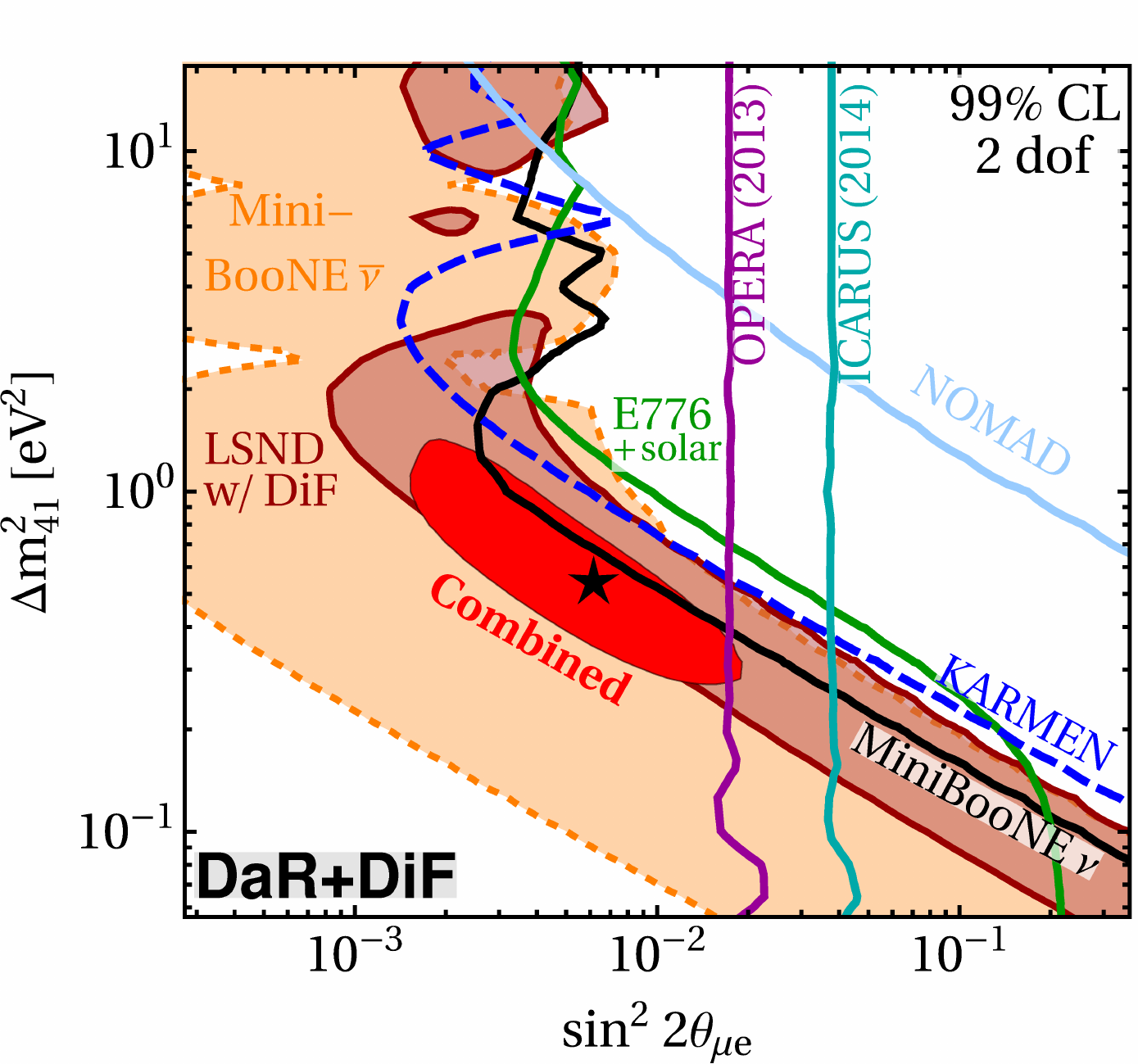}
    \caption{
      \textit{Left}: Status of the 2+2 oscillation scenarios from
      Ref.~{\protect\cite{2+2}}
($\eta_S=\displaystyle\sum_i |U_{is}|^2$ where $i$ runs
over the two massive states mostly relevant for solar neutrino
oscillations).
\textit{Right}: Present status of 3+1 oscillation scenarios from Ref.~{\protect\cite{3+1}}.}
    \label{fig:sterile}%
\end{figure}

The most immediate question as these anomalies were reported was
whether they could all be consistently described in combination with
the rest of the neutrino data if one adds those additional sterile states.
Quantitatively one can start by adding a fourth
massive neutrino state to the spectrum and perform a global analysis
to answer this question. Although the answer is always the same the
way to come about it depends on the way the massive states are ordered.
In brief, there are six possible four-neutrino schemes which can in principle
accommodate the results of solar+KamLAND and atmospheric+LBL neutrino
experiments as well as the SBL 
result. They can be divided in two classes: (2+2) and (3+1). In the
(3+1) schemes, there is a group of three close-by neutrino masses
(as on the 3$\nu$ schemes described in the previous section)  that
is separated from the fourth one by a gap of the order of 1~\eVq,
which is responsible for the SBL oscillations.
In (2+2) schemes, there are two pairs of close masses (one pair responsible
for solar results and the other for atmospheric~\cite{GomezCadenas:1995sj})
separated by the
${\cal O}(\eVq)$ gap. The main difference between these two classes is
the following: if a (2+2)-spectrum is realized in nature, the
transition into the sterile neutrino is a solution of either the solar
or the atmospheric neutrino problem, or the sterile neutrino takes
part in both. This makes this spectrum easier to test as the required
mixing of sterile neutrinos in either solar and/or atmospheric
oscillations will modify their effective matter potential in the Sun
and in the Earth and have observable effects in the data. As described in the
previous section none of those effects were observed and oscillations into
sterile neutrinos did not describe well neither solar nor atmospheric data.
Consequently as soon as the early 2000's 2+2
spectra could be ruled out already beyond 3-4 $\sigma$  as seen
in the left panel in Fig.~\ref{fig:sterile} taken from Ref.\cite{2+2}.

On the contrary, for a (3+1)-spectrum (indeed 3+N), the sterile
neutrino(s) could be only slightly mixed with the active ones and
mainly provide a description of the SBL results. Qualitatively the
constraints on these scenarios come from the tension between the
non-negligible mixing of both $\nu_e$ and $\nu_\mu$ with the additional
massive states required to explain both the LSND/MiniBooNE appearance
results and the $\nu_e$,$\bar\nu_e$ disappearance results from
Gallium and reactor data, with the constraints on the same mixings
from the rest of the data. Again, this is history written as I type
with the upcoming of several reactor experiments designed specifically for
testing these scenarios. The status of the global analysis of the available
data at the time of this talk is
illustrated in the right panel in
Fig.~\ref{fig:sterile} taken from Ref.\cite{3+1} which concluded
that 3+1 scenario is excluded at 4.7$\sigma$ level. Also quoting
from that reference {\it the tension cannot be eliminated by discarding
any individual experiment}. 

\label{sec:sterile}
\section{Epilogue}
Human history is mostly told by the winners. But in neutrino physics, and
in science in general, I would like to think that we can all consider ourselves
winners in one way or another. For me the prize of being an informed witness
of the discovery of beyond the Standard Model Physics has certainly been
worth the effort of countless white nights, stressful last minute talk
updates, and the hundreds of life anecdotes they provoked.

And if that was not enough, it brought me to Paris for this conference
to enjoy the company of great people. Above all the organizers to whom
I remain indebted for their invitation.
\section*{Acknowledgments}
I want to take this opportunity to specially thank Michele Maltoni, my long
time collaborator in the neutrino oscillation analysis. 
This work is supported by USA-NSF grant PHY-1620628, by EU Networks
FP10 ITN ELUSIVES (H2020-MSCA-ITN-2015-674896) and INVISIBLES-PLUS
(H2020-MSCA-RISE-2015-690575), by MINECO grant FPA2016-76005-C2-1-P
and by Maria de Maetzu program grant MDM-2014-0367 of ICCUB.

\end{document}